# An 'Imitation Game' concerning gravitational wave physics

## Harry Collins

**This being Chapter 14 of Collins, Harry (2017, forthcoming) *Gravity's Kiss: The Detection of Gravitational Waves*, Cambridge Mass., MIT Press**

## Chapter 14: The book, the author, the community and expertise.

My contact with the field began in 1972 at the peak of the controversy over whether Joe Weber had detected gravitational waves with his room temperature resonant-bar detectors. The controversy over gravitational waves made up about a quarter of my PhD project in the sociology of science. In those days I conducted my research by going from laboratory to laboratory and interviewing the scientists. I conducted eight interviews in Britain and America in 1972 and a further 14 in Europe and America in 1975/6. The work I did in 1972 gave rise to what became a very well-known paper which secured my career. In 1985 I wrote *Changing Order*, basing it on three of the four sets of interviews I carried out in those years with the gravitational wave research being the central case-study. A much deeper engagement with the field began in in the mid-1990s. Between the mid-1990s and the mid-2000s I spent more time with gravitational wave physicists than with any other group including my fellow sociologists. In those years I travelled to pretty well every conference and workshop the physicists held, often completing more than half-a-dozen flights a year, many of them long-haul. During this period I got to know the community really well and made new friends and acquaintances among them. Crucially, I got to like them and like their project. I felt comfortable among these scientists and privileged to be close to this extraordinary enterprise.

Relevant to this book is the analysis of expertise that has been central to my work for more than ten years. The crucial and most successful concept within that stream of work is 'interactional expertise'.[1] Interactional expertise is best exemplified by my understanding of gravitational-wave physics during the period of my intense involvement; in that time, over coffee, lunch and dinner I would talk gravitational-wave physics with my new friends and acquaintances and do a pretty good job of it even though I was not a physicist myself: I could not do a calculation; I could not contribute to papers; I could not help with building the apparatus. Nevertheless, I thought that someone like myself, with interactional expertise, but, as the language went, without 'contributory expertise', could understand the field to the point of being able to make reasonable technical judgments. I pointed out that peer-reviewers and managers of technological projects are in a not dissimilar position. This led me to try it out and I took part in an 'imitation game' in which a GW physicist asked technical questions – he asked seven in all – of me and another GW physicist. The dialogue – seven questions and seven pairs of answers with identities disguised – was then sent to nine other GW physicists who were asked to identify the participants, knowing one of them was me. Seven said they couldn't work out who was who and two said I was the real physicist. So I passed![2]

Interactional expertise degrades if it is not continually refreshed by constant contact with the changing field of science or technology, or other domain, to which it pertains. Because the intensity of my engagement with GW physics has diminished since the mid-2000s, my expertise has also begun to fade. It was given a couple of boosts as I wrote *Gravity's Ghost* and *Big Dog* but I don't think I ever quite regained the wide level of knowledge I had earlier.

---

[1] See Note XV in Notes and References (below).
[2] See Note XVI in Notes and References (below).

Since the end of the Big Dog analysis another 3 or 4 years has passed during which time I have not been to more than about one meeting a year. I've fallen back further in my understanding of the field, particularly the detailed workings of the instruments. Fortunately the loss of that kind of knowledge has not been fatal; if I was trying to write a second edition of *Gravity's Shadow* it would have been more troublesome.

Luckily this book is not about building detectors but about the way a detection is confirmed once the signals emerge from the completed devices. That means I need a narrow body of understanding which, fortunately, I refined to a pretty high level in writing *Big Dog*.

But to leave it like that would be disingenuous. Therefore I undertook another Imitation Game to see if it would reveal the fading of my expertise. Sathyprakash, Professor of Physics at Cardiff, once more helped by inventing a new set of questions more suited to 2015. He set eight questions as follows.

*Q1. Advanced LIGO and Virgo data contain two signals from identical binary neutron star systems with their coalescence times separated by just 1 second. Explain if you think it would be possible to disentangle the two overlapping signals by matched filtering.*

*Q2. Einstein Telescope is a possible future 3rd generation gravitational wave detector. It is conceived to be an underground detector using cryogenic technology. What sources of noise are being mitigated by going underground and using cryogenics.*

*Q3. An alert from LIGO-Virgo analysis is sent to astronomers a day after a transient event was found. What sort of telescopes (gamma-ray, x-ray, infrared, optical, radio) should astronomers use for follow-up and why?*

*Q4. A continuous gravitational wave signal is observed by a **single** LIGO detector with a high confidence. Explain if you think it will be possible to obtain the sky position of and distance to the source?*

*Q5. It is said that the pulsar timing arrays (PTAs) and laser interferometers are essentially based on the same principle of detection of gravitational waves. What is this principle and how does it apply to PTAs and laser interferometers?*

*Q6. An experimentalist suggests using 10 times greater laser power to improve the sensitivity of an existing interferometer. Assuming that mirrors are able to withstand such an increase in the power and neglecting mirror thermal noise how do you think the strain sensitivity of the detector improves at different frequencies.*

*Q7. LIGO and Virgo take data for 5 years during which there is a galactic supernovae, 200 short, hard gamma-ray bursts and 4 pulsar glitches but fail to detect any signals. Under the circumstances, what are the implications of non-detection?*

*Q8. Two teams, E and N, came up with detector designs both of which had the same distance range for binary coalescences; the E team had better strain sensitivity with a low frequency cutoff of 5 Hz while the N team had relatively poorer strain sensitivity but a slightly lower low frequency cutoff of 1 Hz. Explain if you think they will both be equally good in measuring the parameters of a coalescing binary.*

These questions were answered initially by me and three other gravitational wave physicists. This time I decided to make the test more elaborate and asked other kinds of people to answer too so as to obtain some comparisons. The 'bottom line' is shown in Table 2 below:

|  |  | Who marked the answers? | | | |
| --- | --- | --- | --- | --- | --- |
|  |  | 4 different GW physicists | 2 Savvy social scientists | 2 Social scientists | Harry Collins |
| Who answered the questions? | 3 GW physicists | 27 | 27 | 19 | 23 |
|  | 3 Savvy physicists | 19 | 23 | 17 | 13 |
|  | 2 Savvy soc scientists | 17 | 20 | 19 | 11 |
|  | Harry Collins | 25 | 27 | 20 | 28 |

*Table 2: Overall outcome of new GW Imitation Game*

As can be seen, four groups answered the eight questions – counting Harry Collins as a group. There were 3 GW physicists, there were 3 astronomers/astrophysicists who worked in the same department as the GW physicists (referred to as 'savvy physicists') and 2 social scientists who were acquainted with my work on gravitational waves and with Imitation Game tests of this sort (referred to as the 'savvy social scientists'). Four groups also marked the questions; there were 72 questions and answers in a list randomised at the level of the individual question.[3] Markers have to score the answers according to the following four-point scale giving a maximum of 32 points:

**Knows GW physics:4    Understands:3    Unconvincing:2    Does not know GW physics:1**

Starting with the left hand column we see that Collins did pretty well when marked by the GW physicists – better than I expected – scoring 25 points as opposed to the 27 point mean which the three GW physicists achieved. The effective similarity of these two scores can be seen by noting the wide separation between them and the scores achieved by the non-GW specialist physicists and the social scientists. I believe, however, that my achievement in this test underestimates the erosion of my expertise over the last few years.

This erosion is a bit more evident if we go to the last column. Here I mark all the other answers and mark my own as well. I should add that the marking was done 4 months after I had answered the questions and I recognised only two of my own answers. That I gave myself such a high score is, nevertheless, quite understandable since one is bound to think one's own answers are right. Note also that I mark in a roughly similar way to the GW physicists. Recognising what counts as a right answer is nearly as much of an indicator of understanding as providing the answer in the first place so the fact that there is wide separation between my marks for the GW physicists and those of the physicists and savvy social scientists is a plus for me. The gap between the marks I gave myself and the marks I gave the GW physicists is, however, revealing of the erosion of my expertise. To reiterate, part of this gap is due to the fact that one tends to mark one's own questions well because if one didn't believe them one wouldn't have given them in the first place. Comparing the marks I gave myself and the marks the GW physicists gave me, I think I over-marked myself by two or three points. But I also investigated my low marking of the GW physicists very carefully and it does show that in a couple of cases I did not understand the physicists' answers because of the weakening of my contact with the field. In the case of one answer, I just marked carelessly but I can tie this down to the erosion of my understanding in the case of two other answers. If, over the last few years, I had been hanging around with the GW physicists as intensely as I did in the 1990s and early 2000s I would not have made these mistakes and, being, as it turns out, an overall low marker – we all have different marking tendencies – I would have given the physicists a rounded average mark of 25 instead of 23 plus another point for careless grading of one question. So that is a measure of the erosion of my understanding as measured by this test.

---

[3] Thanks to Luis Galindo for help on this. See Note XVI in in Notes and References (below).

I investigated the answers to Question 2 the most carefully because I thought I had the answer right and the physicists had it wrong. But it turned out, after a lot of inquiries, that the account of gravity gradients that I had in mind, and in my account was the most important reason for building underground, was incorrect. It has been correct a decade earlier when I was more deeply embedded in the science but I had missed the change in understanding that had happened over those ten years. The difference is that for me gravity gradients were mostly about changes in gravitational forces on the mirrors caused by the changes in air density associated with wind and this can be mitigated by going underground to a shallow depth whereas as time went on this effect was found to be small and the serious problem has come to be seen as changes caused by attractions from miniscule ripples in the earth's surface associated with seismic noise; to mitigate these one needs to go much deeper and going that deep is not on many people's minds as there are other ways of compensating for the effect. We can see the difference in these two quotations from the literature from 2000 and 2011 respectively:

> *At such low frequencies environmental effects, and particularly gravity gradients associated with tides and weather variations in the surrounding environment, create perturbations which greatly exceed the desired signal.* (L Ju, D G Blair and C Zhao, 2000, 'Detection of gravitational waves' Rep. Prog. Phys. 63 (2000) 1317–1427, p1326)
>
> *the dominant source of gravity gradients arise from seismic surface waves, where density fluctuations of the Earth's surface are produced near the location of the individual interferometer test masses* (Matthew Pitkin, Stuart Reid, Sheila Rowan, Jim Hough (2011) 'Gravitational Wave Detection by Interferometry (Ground and Space)' http://arxiv.org/pdf/1102.3355.pdf p13)

To repeat, I'd missed this change and that is why I thought the physicists were all wrong and I was right. That is a kind of iconic example of what happens as one's interactional expertise begins to degrade and why it is bound to happen if one is not continually mixing with the community.

I found that apart from the gravity gradients problem, I was also not so good on the relationship between low-frequency and determination of chirp mass and sky position. In retrospect, I can see that if I had not been following things closely since September I would have been out of date on the optimum length of time-slides (Chapter 4) and the status of the idea of a freeze on detector states. But my background knowledge is such that in respect of all of these technical matters, I could have caught up with a short conversation.

Going back to the overall marks, we can see that the non-GW physicists, even when marked by GW-physicists, did not do much better than the savvy social scientists and the separation of GW-physicists plus Collins on the one hand and non-GW-physicists plus social scientists on the other is quite clear. This little tests reveal how specialised science is – there are no 'science experts in white coats'. It is made more striking by the fact that the savvy physicists were pretty savvy. The one who did best wrote to me later:

> For what it's worth I based my answers on what I've learned over the years from seminars by GW people, papers that I've read as a member of peer review and policy committees, and reading semi-popular (Scientific American level) articles. But also the questions about noise and sensitivity made some sense to me as my own field is experimental (electromagnetic) astronomy.

In spite of this, there score was similar to that of the social scientists when marked by GW physicists. Furthermore, the two mean scores, given that there were 8 questions, were 2.4 and 2.1 – both hovering only just above 'unconvincing' on the grading scale. So it was not so much that the savvy social scientists were good, but that the savvy physicists were not very good!

Looking at the middle columns we see that the ability to distinguish between the classes diminishes as the markers grow less expert – this is exactly what we would expect; looking at the third column we see ordinary social scientists mark everyone roughly the same. This show that 'the public' are very easy to fool; all 'experts' look the same to them.

Finally, sometime after the main exercise was completed, I asked two physicists from a university where gravitational waves was not a central concern to mark the test. Their specialties were theoretical optical physics and particle physics respectively. If presented as one of the columns in Table 2 their average rounded scores would be 22, 18, 18, 18, which, in terms of discrimination between the experts and the non-experts, is closer to the non-savvy social scientists than any other group. The two sets of marks, however, were rather different with the second mark being close to random so it seems right to report them both. They were 25, 18, 18, 19 and 19, 18, 19, 17. The first set is the only one to make a marked separation between Collins and the GW physicists; this is puzzling and it may be that stylistic features were playing a role.

Going back to the question of my fading expertise, expertise is multi-dimensional and erosion in respect of another of the less obviously technical dimensions is much harder to rectify – it cannot be rectified by a few minutes of conversation. In our book, *Rethinking Expertise*, we pull out 16 components of expertise or judgements about expertise and even that isn't quite right. In later papers we explain that the main components of technical understanding, interactional and contributory expertise, need further refinement. A vital component of these kinds of expertise is an understanding of the credibility of those who are making expert claims. This is easily seen in the 'meta-expertise' rows of what we call 'The Periodic Table of Expertises' – the rows that cover ability to judge between different experts and enable one to discount certain politicians and salespersons, judge between astronomers and astrologers, judge, or unfortunately fail to judge, between doctors and vaccine-scare mongers, and, most usefully, judge between scientists paid by the tobacco and oil companies to deliver certain results favourable to them and those who are driven by scientific values. But that kind of meta-judgement also makes up a component of the 'purely' technical expertises.

Our standard and oft-repeated example is, once more, taken from gravitational wave physics. In 1996 Joe Weber published a paper claiming to have a found a correlation between gamma-ray bursts and the gravitational waves he had found in earlier years. I went around the community asking scientists what they made of this paper. I discovered that I was the only person to have read it. The 'technical' judgment being made was that Weber's credibility had now fallen so low that this, in spite of its perfectly respectable appearance, was a 'non-paper'. arXiv, the physics preprint server, uses a computer algorithm to screen submissions. In 2015 we discovered that this paper, albeit twenty years old, still passed arXiv's screening process without problem. Indeed, to outsiders the paper has all the appearance of a potential Nobel-prize-winning contribution to physics.[4]

In a paper published in 2011we call the kind of expertise needed to make the judgment that Weber's 1996 paper should be ignored, 'Domain-Specific Discrimination' and define it as, 'the ''non-technical'' expertise used by technical experts to judge their fellow experts'.[5] Apart from what has been shown by my inadequate marking of the GW question test, I have lost the ability to make this kind of judgment in respect to a good proportion of the community as result of my years of distance from the field; this is where I *feel* the loss of expertise most strongly; given what I already know, the technical losses can be patched with a few email inquiries or telephone discussions but you cannot get to know the competencies and biases of a community like this – that takes years. Over the last few years many new people have entered the field or shifted from relative obscurity to positions in which they are making significant contributions.

Even when I wrote *Gravity's Ghost* and *Big Dog*, both of which depended heavily on my perusal of the email traffic, I knew pretty well all the people behind the emails and I knew their political stakes in the matter – I could see where people were coming from and assess their contributions accordingly. This time around I do not know what half the contributions mean in the sense of *domain-specific discrimination* – I do not grasp how seriously to take what is being said because I do not know the person saying it and *why* they are saying it. Thus, I went through the 104 emails that I saved rather than deleted in on the first Monday and Tuesday of the appearance of G150914 and found there were 45 separate emailers of whom I roughly knew only 20. A few years ago those numbers would have been, perhaps, 40 and 5. That makes a lot of difference to how one understands a field.

Here is an example from Tuesday 15th when I remark to one of the scientists:

> I found the argument that YYYY because of ZZZZ very weird

to which came the response

> A very weird argument indeed from XXXX, although perhaps not so weird considering it was from XXXX ;-)

On the other hand, in another conversation on 28th I am told:

> I think before we're done we're are going to have to understand whether there is any credibility (44.42) to that … and I think that's going to be a struggle because yyyy is a really smart guy and he's pretty self-confident and he will say he believes it and people have enough respect for him that they will not blow him off so I don't know how we're going to resolve that. [Interview with Peter 28 Sept]

As it happens I know the person referred to in the second comment well enough to understand and appreciate what is being said. But I don't the person who is the subject of the first comment and in general this is the kind of thing that I am finding much harder to work out for myself in respect of GW50914 than Big Dog – I am, to repeat, losing Domain Specific Discrimination (DSD). It is easy to see how DSD affects face-to-face communication and in another place I show how lack of familiarity causes me to make a mistake over the body language of someone with whom I am heatedly discussing an issue, but what has become clear to me since the event is that the same applies to the interpretation of emails.[6] The written word is not just written: its meaning rests on a raft of previous social interactions. This is a very important point in these days when communication among the younger generation appears to rest more-and-more on social media.

---

[4] The paper is Weber and Radak, 1996. See Note XVII in in Notes and References (below).
[5] See Note XVII in in Notes and References (below).

**Notes and References**

*Notes*

**XV**  *Interactional Expertise*: The idea of Interactional Expertise brings language into central focus. It is argued that by spending enough time taking part in the spoken discourse of a specialist group – by acquiring their 'practice language' (Collins, 2011) – one can learn to understand their world of practice without taking part in the practice itself. This idea flies in the face of a long tradition in philosophy which stresses the central importance of practice to understanding.

The origin of the argument is debates in artificial intelligence (AI). The influential philosopher Hubert Dreyfus, drawing on the philosophers Heidegger and Merleau-Ponty, argued in a famous (1967) paper and subsequent books (1972, 1992) that computers would never have human-like intelligence unless they had bodies with which they could move around and experience the world in the ways that humans experience it. AI enthusiast Doug Lenat argued that this must be wrong because of the capacities of persons who did not possess ordinary human-like bodies. His example was 'Madeleine'; Madeleine was severely disabled from birth yet was entirely fluent with her fluency attained through conversation not physical interaction (Sacks, 2011). The response to this by philosophers (eg Selinger; 2003; Selinger at al, 2007) was that Madeleine had a body with front, back, etc and could work from this. But this answer allows that even if we need some kind of vestigial body to have intelligence (the argument seems doubtful but let us allow it), we don't need much of a body to learn everything practical we need to learn; we can learn from conversation.

Collins (eg 1996) tried to resolve the problem by splitting it into two. Human societies, or specialist groups within those societies, would not be human-like unless they had bodies, but individuals can learn from those societies without sharing their bodily form.

> Wittgenstein said that if a lion could speak we would not understand it. The reason we would not understand it is that the world of a talking lion - its `form of life' - would be different from ours … lions would not have chairs in their language in the way we do because lions' knees do not bend as ours do, nor do lions `write, go to conferences or give lectures'.  ... But this does not mean that every entity that can recognise a chair has to be able to sit on one.  That confuses the capabilities of an individual with the form of life of the social group in which that individual is embedded.  Entities that can recognise chairs have only to share the form of life of those who can sit down.  We would not understand what a talking lion said to us, not because it had a lion-like body, but because the large majority of its friends and acquaintances had lion-like bodies and lion-like interests. In principle, if one could find a lion cub that had the potential to have conversations, one could bring it up in human society to speak about chairs as we do in spite of its funny legs. It would learn to recognise chairs as it learned to speak our language. This is how the Madeleine case is to be understood; Madeleine has undergone linguistic socialization. In sum, the shape of the bodies of the members of a social collectivity and the situations in which they find themselves give rise to their form of life. Collectivities whose members have different bodies and encounter different situations develop different forms of life. But given the capacity for linguistic socialisation, an individual can come to share a form of life without having a body or the experience of physical situations which correspond to that form of life.[7]

---

[6] The other place is my book 'Artifictional Intelligence' (wherever this stands at the time this ms is being finalised).XXXX

To put this in terms of more familiar examples, one cannot have a tennis language to learn from unless there are groups of humans with the physical ability to play tennis but one can learn what it is to play tennis and, in principle, all its practical nuances, just by talking to tennis players. The same, of course, goes for gravitational wave physics.

It is not easy to acquire interactional expertise – it takes a long time – but once acquired it is much more than the ability to 'talk the talk'. It is better described as being able to 'walk the talk'. It has much in common with the kinds of knowledge managers of technical projects possess (Collins and Sanders, 2007). The idea of interactional expertise seems necessary if we are to understand many features of the way societies work, how they support the division of labour in technical specialties and the way sub-groups interact with society as a whole. A useful discussion of the concept in the context of a classification of different types of expertise is Collins and Evans, 2007, while the most recent and complete review is Collins and Evans 2015.

As explained in the text, it is the idea of interactional expertise that makes it possible to contemplate an outsider without a physics degree undertaking a project like this one. It is also the idea of interactional expertise that gives impetus to Imitation Game exercises such as that discussed in Chapter 14 and Note XVI.

**XVI** *Imitation Games and Turing Tests*: As mentioned in Note XV, Imitation Games are associated with interactional expertise. Imitation Games were the precursor to the Turing Test. Turing based his test on a parlour game in which hidden men and women pretended to be each other while responding to written questions from a judge. Turing believed that a computer should be called 'intelligent' if, say, it was as good at pretending to be a woman as was a man pretending to be a woman; to reiterate, what was supposed to be indistinguishable according to the original description of the test (Turing 1950) was a computer pretending to be a woman versus a man pretending to be a woman (or vice versa), not a computer pretending to be a human compared to a human – the way so-called Turing Tests are conducted these days.

We use Imitation Games, that is, Turing Tests with humans, to test for the possession of interactional expertise. A classic example is an experiment in which blind persons pretending to be sighted are compared to sighted persons with sighted persons asking the question. The questioning can be mediated via interlinked computers. This configuration is compared to one where sighted persons pretend to be blind while being compared to blind persons with blind persons asking the questions.  In these pairs of tests, the blind persons pretending to be sighted do much better that the reverse because blind persons are immersed in the spoken discourse of the sighted whereas the sighted are not, generally, immersed in the discourse of the blind.  Thus blind persons have many opportunities to acquire interactional expertise in the world of the sighted.  We have run such tests at various scales and on various topics (Collins and Evans, 2014).

---

[7] From a review of Hubert Dreyfus's book, *What Computers Still Can't Do*, (Collins 1996), which was published in the journal *Artificial Intelligence.*

The original gravitational wave imitation game test was run over email with a gravitational wave physicist setting seven questions which were answered by Collins along with another gravitational wave physicist.  The completed dialogues were sent to nine other gravitational wave physicists who were asked: 'Which is the real gravitational wave physicist and which is Harry Collins?'  Seven of these said they could not tell the difference and two said Collins was the real gravitational wave physicist.  An account was written up as a news item in *Nature* (Giles, 2006).  It is important to understand, however, that contrary to what the Nature story can be read to imply, this exercise was not a hoax but a demonstration of genuine understanding as exhibited by a display of interactional expertise.  A more elaborated version of this test applied to Collins's slightly eroded level of interactional expertise is described in Chapter 14.  Interactional expertise is, of course, closely related to the concept of tacit knowledge – see Notes VIII and XV.  Collins, 2010, is an analysis of tacit knowledge.

**XVII** *Domain discrimination, specialist expertise, and the fringes of science*:  As explained (p237xxx), the sociologist must studiously avoid short-circuiting the process of inquiry into the social factors that feed into scientific belief.  This implies that truth, or rationality, or similar cannot be allowed to be part of the explanation of why something came to be believed.  It thus becomes very difficult for the sociologist to distinguish between mainstream science and what we can call 'fringe science'.  There is a large, organised, fringe with its own journals and its own annual conferences (Collins, Bartlett and Galindo, 2016 arXiv XXXX).  One concern of many fringe scientists is a rejection of the theory of relativity; some claim the theory is a massive conspiracy, even, given Einstein's support for Israel, a Jewish conspiracy!  Whatever, rejection of relativity implies rejection of The Event.  Chapter 11 gives some examples of such rejections.
The self-denying ordinance of the sociologist makes the partition of science into mainstream and fringe a much more interesting problem that it is when analysts simply allow themselves the luxury of being parasites on the opinions of the scientists themselves – 'scientists say relativity is right so people who do not believe in it are irrational'.  Instead one must try to find sociological demarcation criteria, a much more demanding task.  It is an important task, not so much for the future of scientific knowledge where what counts as the truth will emerge over the decades, but for the decisions that have to be made by policy-makers.  For example, in 1992 (*Gravity's Shadow* p 361) Joe Weber wrote to his Congressional representatives indicating that his new theory of the sensitivity of resonant bars meant that the much more expensive interferometers were a waste of money.  This presented no problem to the mainstream community but how was an outside decision-maker to judge the claim?  The only answer seems to be that that in a democracy decision-makers will have to base their decisions on the opinions of the mainstream institutions and that is why demarcation criteria, that are more than the opinions of scientists, are needed.  Strangely enough arXiv has a similar problem in that it receives many submissions from fringe papers and the sheer logistics of the operation demands some automation.  In collaboration with Paul Ginsparg, the founder of arXiv, (Collins, Galindo and Ginsparg, 2016) we have shown that the automated methods used by arXiv, though they represent the state of the art, do not recognise papers, such as Weber and Radak, 1996, as anything out of the ordinary.

What is needed in those cases is what we call 'Domain Specific Discrimination' or 'Domain Discrimination' for short (Collins and Weinel, 2011, p407).

In Collins, Bartlett and Galindo 2016, we establish a series of characterisations of the fringe. Indicative is the difference between the fringe and the mainstream in respect of what Thomas Kuhn (1959, 1977) called 'the essential tension'. The essential tension is that between the need to preserve the right of the individual to make novel claims setting him or herself outside of the consensus and the need to accept a degree of regulation of scientific thinking and acting if science is to move forward. In the normal way, science is always balancing these two needs. We find that in the fringe the balance shifts markedly to the side of the individual with consensus being thought dull or suspiciously authoritarian; this shift in the balance is even visible in fringe scientific conferences with each delegate's pet theory being given space to be expressed resulting in a general lack of organisation. In this book, of course (Chapter 11), given the uniform lack of criticism among the mainstream, we find ourselves drawing on the fringe to provide criticism of the first detection of gravitational waves; it is the fringe, in refusing to accept the social consensus, that allows us to see the extent to which acceptance of The Event is a matter of social consensus!

### *Four previous books touching on gravitational waves*

Collins, Harry (1985) *Changing Order: Replication and Induction in Scientific Practice*, Beverley Hills & London: Sage. [2nd edition 1992, Chicago: University of Chicago Press]

Collins, Harry (2004) *Gravity's Shadow: The Search for Gravitational Waves*, Chicago: University of Chicago Press

Collins, Harry (2011), *Gravity's Ghost: Scientific Discovery in the Twenty-First Century*, Chicago: University of Chicago Press

Collins, Harry (2013), *Gravity's Ghost and Big Dog: Scientific Discovery and Social Analysis in the Twenty-First Century*, Chicago: University of Chicago Press [includes a reprint of *Gravity's Ghost*]

-------------------------------------------------------------------------------

### *Other references*

Collins, Harry, 2011. `Language and Practice' *Social Studies of Science,* 41, 2, 271-300

Collins, Harry. 1996. "Embedded or Embodied? A Review of Hubert Dreyfus' What Computers Still Can't Do." *Artificial Intelligence* 80(1):99–117.Collins, Harry, 2011. `Language and Practice' *Social Studies of Science,* 41, 2, 271-300

Collins, Harry, Bartlett, Andrew and Reyes-Galindo, Luis, 2016, 'The Ecology of Fringe Science and its Bearing on Policy', (http://arxiv.org/abs/1606.05786)

Collins, Harry, and Evans, Robert, 2014, 'Quantifying the Tacit: The Imitation Game and Social Fluency' *Sociology* 48, 1, 3-19